\documentclass[%
 reprint,
showpacs,preprintnumbers,
 amsmath,amssymb,
 aps,
prl,
floatfix,
]{revtex4-1}

\usepackage{graphicx}
\usepackage{dcolumn}
\usepackage{bm}
\usepackage{hyperref}

\begin{document}

\preprint{APS/123-QED}

\title{Friction-mediated flow and jamming in a two-dimensional silo
  with two exit orifices}

\author{Ashish V. Orpe} \email{av.orpe@ncl.res.in}
\affiliation{Chemical Engineering and Process Development Division,
  CSIR-National Chemical Laboratory, Pune 411008 India}
\affiliation{Academy of Scientific and Innovative Research (AcSIR),
  Ghaziabad 201002 India}
\author{Pankaj Doshi}
\email{pankaj.doshi@pfizer.com}\affiliation{Pfizer Inc., Groton,
  Connecticut, 06340 USA}

\date{\today}

\begin{abstract}
  We show that the interparticle friction coefficient significantly
  influences the flow and jamming behavior of granular materials
  exiting through the orifice of a two-dimensional silo in the
  presence of another orifice located in its vicinity. The
  fluctuations emanating from a continuous flow through a larger
  orifice results in an intermittent flow through the smaller orifice
  consisting of sequential jamming and flowing events. The mean time
  duration of jammed and flow events, respectively, increase and
  decrease monotonically with increasing interparticle friction
  coefficient. The frequency of unjamming instances ($n_{u}$),
  however, shows a nonmonotonic behavior comprising an increase
  followed by a decrease with increasing friction coefficient. The
  decrease on either side of the maximum, then, represents a system
  moving progressively towards a permanently jammed or a permanently
  flowing state.  The overall behavior shows a systematic dependence
  on the interorifice distance which determines the strength of the
  fluctuations reaching the smaller orifice leading to unjamming
 instances. The probability distributions of jamming and
  flowing times are nearly similar for different combinations of
  friction coefficients and interorifice distances studied and,
  respectively, exhibit exponential and power-law tails.
\end{abstract}

\pacs{45.70.Mg,47.57.Gc}
\maketitle

\section{\label{intro}Introduction}

A dry granular material exiting from an hopper or a silo can jam
abruptly and quite unpredictably on its
own~\cite{to01,tang11,tewari13,zuriguel14b}. On the contrary, the same
jammed orifice requires forced intervention in some form to unjam or
reinitiate the flow.  The occurrence of the former is due to a stable
arch formed at the exit and is dependent on the critical ratio of the
size of particles to the orifice size, which is well defined for a
three-dimensional hopper~\cite{zuriguel03}, but not necessarily for a
two-dimensional hopper~\cite{janda08}. The latter phenomena occurs due
to independent forcing of some form, primarily directed toward
breaking of the arch, viz., impinging of an air jet through the
orifice or vibration of the silo or a hopper~\cite{janda09b} or the
presence of flow through a nearby additional
orifice~\cite{kunte14,mondal14}. While jet impinging or system
vibration form external means of forcing, the presence of another
orifice represents internal forcing, i.e., the inherent flow
characteristics of the silo, in the form of velocity fluctuations, fed
onto itself to cause unjamming which can also lead to improved
mixing~\cite{kamath14}. 

The continual presence of such an independent parallel forcing results
in an intermittent flow through the orifice (flow followed by jamming
followed by flow and so on), which can vary from a continuous flow
regime to a permanently jammed regime dependent on the propensity of
the forcing, i.e., vibration intensity~\cite{janda09b} or the distance
between two orifices~\cite{kunte14}. Within the intermittent flow
regime, the distribution of the times during which orifice is flowing
exhibits an exponential tail, while those corresponding to jammed
state exhibits a power-law tail~\cite{janda09b}. The former represents
the characteristic of a random behavior and is also observed during
the flow from an orifice even in the absence of any independent
forcing~\cite{zuriguel11,kunte14}. The latter behavior is shown to
comprise two different regimes depending on the value of the
power-law exponent~\cite{janda09b}. For values of exponents $2$ and
lower, the distributions comprise jammed events of increasingly
longer durations separating two consecutive flowing events. The
average jamming time is ill defined and increases with increase in the
total experimental duration, eventually diverging over very long
durations suggestive of an overall jammed state. However,
the progressively increasing exponent value above $2$ leads to an
overall flowing state with well defined mean jamming time. The
exponent value of $2$, thus, corresponds to a jamming to flowing
transition. 

Interestingly, this value of the exponent of $2$ is quite insensitive
to the type of independent forcing and has been shown to be valid for
a variety of systems ranging from those occurring naturally
(e.g., movement of a crowd of pedestrians or animals through a narrow
exit) or artificially (e.g., motion of an assembly of granular or
colloidal particles through an orifice
)~\cite{zuriguel14a,zuriguel17,hidalgo18}. The overall behavior can be
qualitatively predicted using an empirical model based on the Langevin
equation with vibrations mimicking the thermal
fluctuations~\cite{nicolas18}.  The different systems, albeit showing
similar universal behavior for jamming-to-flow transition, can be
thought of possessing different friction coefficient between
constituent entities. This, apparent effective friction coefficient,
can owe its origin to different material
characteristics. Experimentally, it has been shown recently that the
(continuous) flow rate of granular material through a two orifice silo
exhibits a qualitative change in its dependence on interorifice
distance with increase in interparticle friction
coefficient~\cite{fullard19}. The continuous flow of such granular
material draining through two orifices, located far apart from each
other but at various distances from the side walls has been predicted
very well using kinematic theory based arguments~\cite{maiti17}

In this work, we strive to explore the jamming and flowing phenomena
through the orifice of a silo for varying interparticle friction
coefficients and for the independent forcing occurring through the
second continuously flowing orifice using discrete element method
(DEM) simulations. This forcing will, thus, depend on the flow through
the second orifice, its proximity to the jammed or flowing orifice, and the
transmission of this forcing though the bed of grains. In the next
section, we describe the system and simulation details, followed by
the results comprising primary causes of unjamming phenomena and
the relevant characteristics of the jamming and flowing behavior.

\section{\label{method}Methodology}

The DEM simulations methodology employed and the system geometry is
nearly identical to that used in previous
work~\cite{kunte14,rycroft09} and we provide only relevant details
over here. The simulations, carried out using the Large
Atomic/Molecular Massively Parallel Simulator (LAMMPS), employ a
Hookean force between two contacting particles which consists of a
normal component ($\bm{F_{n}}$) and a tangential component
($\bm{F_{t}}$). Each of this force has two terms, a contact force and
a damping force given as~\cite{lammps}
\begin{equation}
  \bm{F_{n}} = \left (k_{n} \delta \bm{n} - \frac{\gamma_{n}
      \bm{v}_{n}}{2} \right),
\end{equation}
\begin{equation}
  \bm{F_{t}} = -\left (k_{t} \Delta \bm{s}_{t} + \frac{\gamma_{t}
      \bm{v}_{t}}{2} \right),
\end{equation}
where, $\bm{n}$ is the unit vector along the line connecting centers
of two particles, and $\bm{v}_{t}$ and $\bm{v}_{n}$ are, respectively, the
tangential and normal components of particle velocities. The normal
damping term ($\gamma_{n}$) is chosen as $50 \sqrt{g/d}$, while the
tangential damping term ($\gamma_{t}$) is set as $\gamma_{n}/2$. The
normal elastic constant ($k_n$) is chosen as $2 \times 10^6 mg/d$
while the tangential elastic constant $(k_{t})$ is set as
$2/7~k_{n}$. The elastic constants represent a stiffer particle in
accordance with previous studies~\cite{landry03,rycroft09}.
$\Delta \bm{s}_{t}$ is the tangential displacement between two
particles to satisfy the Coulomb yield criterion given by
$\bm{F_{t}} = \mu \bm{F_{n}}$, where $\mu$ is the friction
coefficient, varied from $0.001$ to $0.5$.  In the above expressions,
$d$ is the particle diameter, $g$ represents gravity acting in
downward direction and the particles have unit density which yields
the mass ($m_{p}$) of the particle as $4 \pi (d/2)^{3} m/3$ with a
natural mass unit $m$. The natural time unit $\tau$ is given as
$\sqrt{d/g}$ and the integration time step used in the simulation is
$\delta t = 2.5 \times 10^{-5}$.

\begin{figure}
  \includegraphics[width=0.95\linewidth]{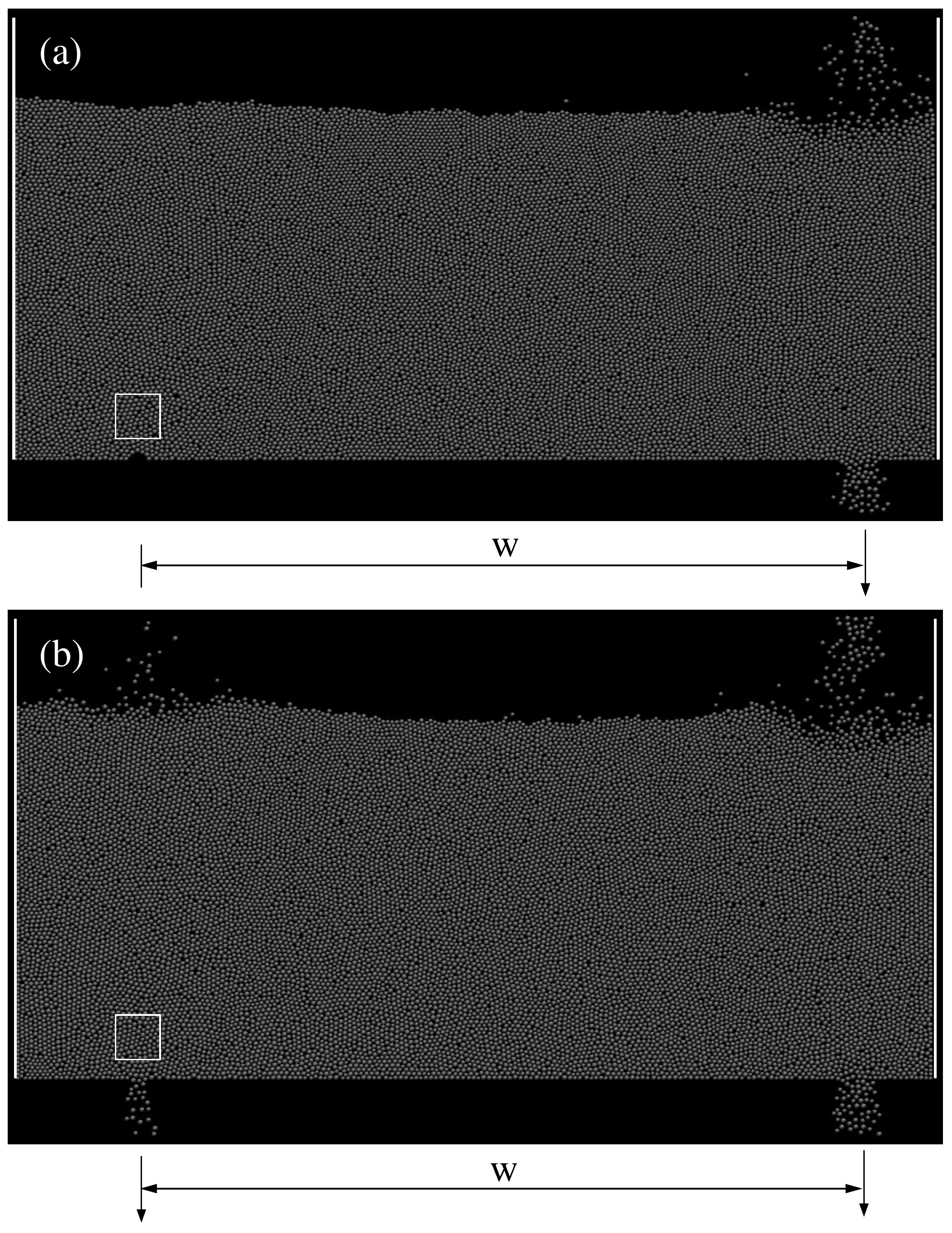}
  \caption{Sample snapshots of the flow occurrence in a two-orifice
    silo for $w = 140d$. Flow occurs continuously through right
    orifice of width $8d$ while the left orifice (width $4d$) is
    jammed as shown in panel (a). The flow reinitialization occurring
    spontaneously through the left orifice at a later time is shown in
    panel (b). The white box ($10d \times 10d$) represents the region
    over which the time dependent mean velocity and average rms
    velocity is calculated. The white vertical lines represent the
    flat side walls. The pouring near the free surface represents the
    granular recirculation (see text for more details).}
  \label{sim-schem}
\end{figure}

A two-dimensional rectangular, flat bottomed silo geometry of thickness
$1d$ is employed in this work. The width of the silo is
specified in terms of $d$ and the silo is filled upto an approximate
height of $80-90d$, with $d$ being the mean particle diameter
with a polydispersity of $15$ \%. The bottom surface is created using
smaller particles ($0.1d$) to mimic a smooth wall, which are kept
frozen during the entire simulation run having zero translational and
angular velocities. The flat side walls are created using the in-built
function in LAMMPS. The friction coefficient between the flowing
particles and both, the side and bottom, walls is maintained same as
the interparticle coefficient. The simulation has two orifices of
fixed widths ($D_{1} = 8d$ and $D_{2} = 4d$) separated by a distance
$w$.  The size of the larger orifice ($D_{1}$) was so chosen as to
allow for continuous flow of particles throughout the simulation
run. The silo width is maintained large enough for all $w$ to
prevent any sidewall effects. The silo is initially filled using the
sedimentation method as suggested previously~\cite{landry03} in which
a dilute packing of nonoverlapping particles is created in a
simulation box and allowed to settle under the influence of
gravity. The simulation is run for a significant time so that the
kinetic energy per particle is less than $10^{-8} mgd$ resulting into
a quiescent packing of desired fill height in the silo which
defines the initial state.

Both the orifices are opened simultaneously to initiate the flow. The
flow through larger orifice occurs continuously without any
interruption, while that through the smaller orifice shows
intermittent flow: several successive sequences of flow and
nonflow. Fig.~\ref{sim-schem}(a) shows a sample snapshot of particles
flowing through larger (right) orifice while the smaller (left)
orifice is jammed. After a while, the flow restarts through the jammed
orifice with continual flow through the larger orifice as shown in
Fig.~\ref{sim-schem}(b).  The fill height ($80-90$d) is maintained
constant by re-pouring the particles which exit the orifice,
from a fixed distance above the free surface at the same horizontal
location where they exited from the silo (see
Fig.~\ref{sim-schem}). Every simulation is executed for $500$ million
timesteps, which provides several jamming-unjamming sequences good
enough to obtain meaningful averages. The total simulation time
corresponds to that required by the particles to traverse the entire
silo height atleast $200$ times. The snapshots of particle positions
within the silo are saved at intervals of $0.25 \tau$.

\begin{figure}
  \includegraphics[width=1.0\linewidth]{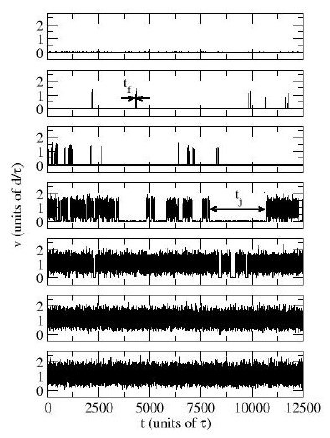}
  \caption{Variation of mean velocity ($v$) of the flow with time
    ($t$) occurring through the left orifice shown in
    Fig.~\ref{sim-schem} for $w = 140d$. The values of friction
    coefficient ($\mu$) vary from top panel to bottom panel as $0.5$,
    $0.35$, $0.1$, $0.05$, $0.025$, $0.01$ and $0.001$,
    respectively. The time duration of flowing and jammed states are,
    respectively, represented by $t_{f}$ and $t_{j}$.}
  \label{flow-time}
\end{figure}

\section{\label{results}Results and Discussion}

The occurrence of flow through the smaller orifice ($D_{2} = 4d$) can
be represented in terms of instantaneous mean velocity ($v$)
calculated inside the silo in a region $10d \times 10d$ centered at a
position exactly $10d$ above the orifice and along the centerline
through the orifice. The region is shown as white box in both the
panels of Fig.~\ref{sim-schem}. The mean velocity is defined as
$v = \sqrt{\langle c_{x} \rangle^{2}+\langle c_{y}
  \rangle^{2}}$. Here, $c_{x}$ and $c_{y}$ are, respectively, the
instantaneous horizontal and vertical velocity components of every
particle obtained from the displacements between two successive
snapshots and $\langle . \rangle$ represents a spatial average over a
region $10d \times 10$d as defined above.  The variation of mean
velocity with simulation time is shown in Fig.~\ref{flow-time} for one
particular interorifice distance ($w = 140d$) and varying
interparticle friction coefficients. A schematic similar to
Fig.~\ref{flow-time} has been presented previously~\cite{zuriguel17}
for a single orifice silo vibrated continuously at different
intensities.  The time $t_{j}$, depicted in the fourth panel, is
defined as the time during which the left orifice remains jammed.
Similarly, $t_{f}$, shown in second panel is defined as the time
during which the flow occurs through the left orifice before it gets
jammed. 
 
Several features, corresponding to jamming and flow occurrences in the
silo, are evident from Fig.~\ref{flow-time}. The mean velocity value
is zero at all times for $\mu = 0.5$ (top panel), which corresponds to
the orifice remaining jammed at all times. A slight reduction in the
interparticle friction coefficient ($\mu = 0.35$) shows few
occurrences of sudden rise in the mean velocity, but only for a very
brief time, followed by rapid decrease to zero velocity. These are
seen as spikes emanating from zero velocity line as shown in second
panel in Fig.~\ref{flow-time}. The orifice, thus, remains jammed
throughout with occasional spurts of flow for a brief period of
time. The mean, $\langle t_{f} \rangle$, is quite low for this case,
while $\langle t_{j} \rangle$ is quite high. Here, $\langle . \rangle$
represents average over entire simulation run.  The instances of flow
re-initiation increase continuously with decreasing friction
coefficient. Further, the duration of the flow following the unjamming
event also increases continuously as evident from contiguous clusters
of spikes. With a significant decrease in the friction coefficient
($\mu = 0.025$, panel five in Fig.~\ref{flow-time}), the situation
gets reversed with the flow now occurring almost at all times with
sudden occasional dips in the velocity to zero, for a brief period of
time. In this case, the mean, $\langle t_{f} \rangle$, is quite high,
while $\langle t_{j} \rangle$ is quite low. This scenario represents
occasional jamming of orifice in an otherwise continuous
flow. Decreasing the value of $\mu$ further eliminates these
occasional jamming events as well leading to a continuous flow
throughout (i.e., diverging $\langle t_{f} \rangle$).  This behavior is
exactly opposite to that observed for $\mu = 0.5$ (top panel), for
which $\langle t_{j} \rangle$ diverges.  Apparently, the state of the
orifice shows a step change from presence of a continuous zero
velocity to a continuous nonzero velocity of an approximate magnitude
of $1.2 d/\tau$. This nonzero velocity shows fluctuations about the
mean, which are perhaps due to lack of smooth flow, possible only
through an orifice of larger size, for instance, $D_{1}$. The overall
behavior from one state to other through a transition is clearly due
to the presence of continuous, smooth, steady flow occurring through
right orifice ($D_{1} = 8d$), in the absence of which, the jammed
(left) orifice will not be able to unjam again~\cite{kunte14}. The
probable cause of this time dependent, friction-dependent and
interorifice-dependent jamming-unjamming behavior is discussed next.

\begin{figure}
  \includegraphics[width=1.0\linewidth]{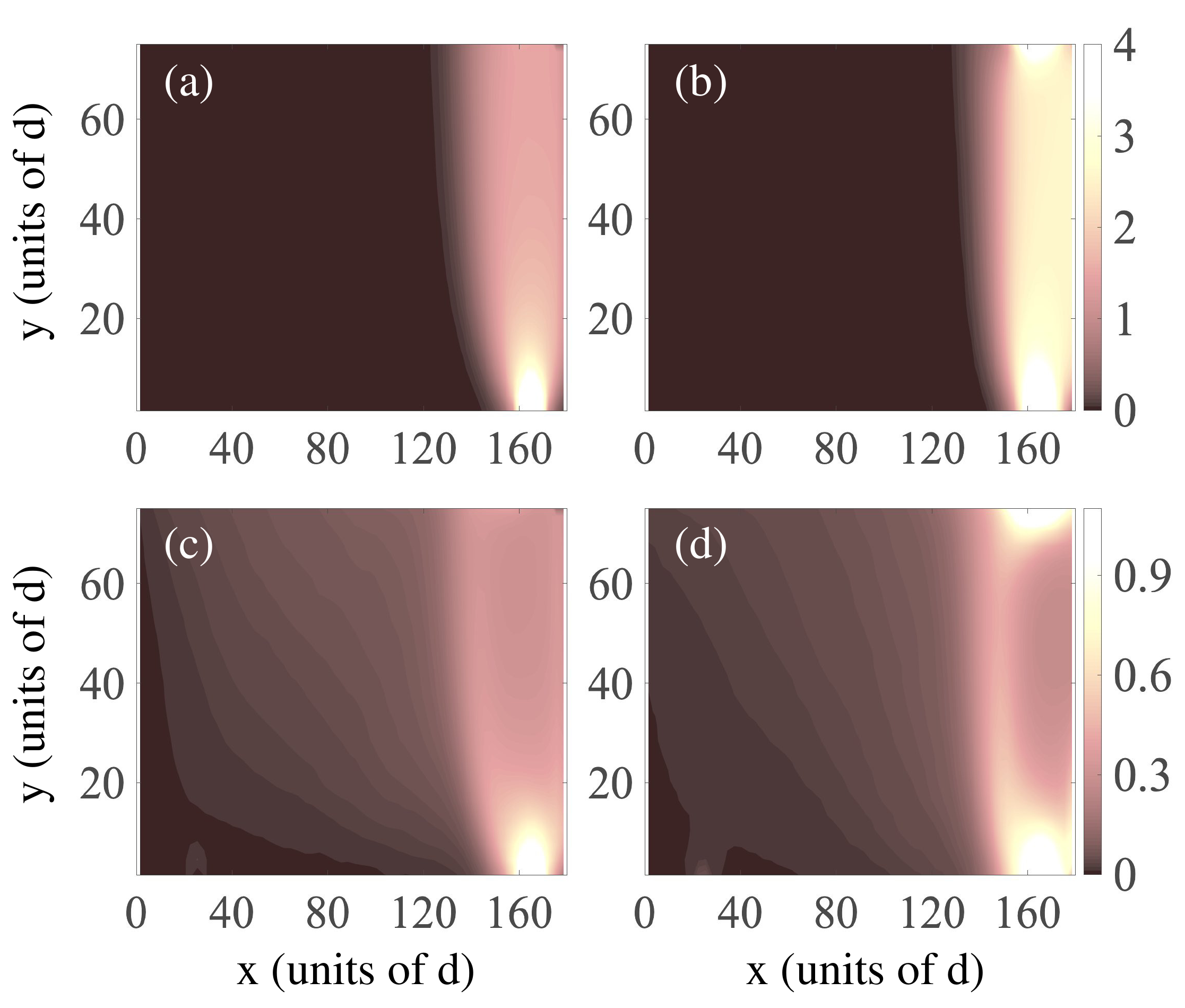}
  \caption{Contour plot of velocities in the silo for an interorifice
    distance of $w = 140d$ and two different friction
    coefficients. The contours are obtained as averages over all those
    times when the left orifice at $x = 25d$ remains jammed while the
    flow occurs continuously through the right orifice at $x = 165d$.
    Mean velocity field for (a) $\mu = 0.35$ and (b) $\mu = 0.05$. RMS
    velocity field for (c) $\mu = 0.35$ and (d) $\mu = 0.05$.  The
    scale (color bar) common to each row is shown on the extreme right.}
  \label{velocity-fields}
\end{figure}

The velocity contours across the entire silo are obtained for all
those times when the left orifice remains jammed.  The mean velocity
($v$) at different locations is calculated using the expression
mentioned above while the fluctuations of mean velocity are measured
in terms of root mean squared (rms) velocity which is defined as
$u =\sqrt{[\langle c_{x}^{2} \rangle - \langle c_{x} \rangle
  ^{2}]+[\langle c_{y}^{2} \rangle - \langle c_{y} \rangle
  ^{2}]}$. For both quantities, $\langle \rangle$ represents the
spatial average over a $3d \times 3d$ region and the temporal average
over all the time instants whenever the left orifice is in a jammed
state. The spatial region for averaging is chosen large enough to get
better statistical averages, but is small enough to reasonably
represent the contour map. The contour map for mean and rms velocity
in the silo for an interorifice distance of $140$d and for two
different friction co-efficients is shown in
Fig.~\ref{velocity-fields}. Since the contours are obtained only for
those times when the left orifice remains jammed, the observed spatial
variation in the velocity magnitudes is the outcome of the continuous
flow occurring through the right orifice located at $x = 165d$. The
mean velocity has a nonzero magnitude only in a small vertical band
($x > 130d$) stretching from the orifice to the free
surface. Elsewhere the mean velocity is close to zero. The unjamming
of left orifice, located at $x = 25d$ and requiring slightest of
relative motion between particles forming the arch, in that case does
not seem to arise due to mean velocity field.  Note that the change in
the friction coefficient (even by an order of magnitude difference)
does not seem to affect the spatial variation of mean velocity
[Figs.~\ref{velocity-fields}(a) and \ref{velocity-fields}(b)].

The rms velocity, however, shows a much broader spatial variation in
the silo and seems to exhibit dependence on the friction coefficient
[Figs.~\ref{velocity-fields}(c) and \ref{velocity-fields}(d)]. For
both, $\mu = 0.05$ and 
$\mu = 0.35$, the fluctuations are observed to be present almost
everywhere in the system, which are expected to cause the relative
motion of the particles in the arch leading to unjamming. The
fluctuations do, however, show an increased spatial coverage for
$\mu = 0.05$ when compared to those observed for $\mu = 0.35$, and
they extend upto the jammed orifice in the former case.  Increased
tangential damping during particle-particle contacts is expected for
increased friction coefficient ($\mu = 0.35$) thereby weakening the
fluctuations reaching the jammed orifice which are, thus, not evident
in the figure scale. These weakening fluctuations may cause cumulative
relative motion of particles in the arch, but over a much longer time
duration, thereby increasing the duration of the jammed events (see
Fig.~\ref{flow-time}, second panel), consequently higher
$\langle t_{j} \rangle$. In the similar vein, the lower value of
$\mu = 0.05$ will cause relatively stronger fluctuations to be present
in the vicinity of the orifice, thereby causing the orifice to remain
jammed for a smaller duration of time (see Fig.~\ref{flow-time},
fourth panel), consequently, lower $\langle t_{j} \rangle$.  In the
event of a very high friction coefficient ($\mu = 0.5$), the
fluctuations reaching the left orifice are not of significant
magnitude to cause unjamming even once over the entire simulation run
(Fig.~\ref{flow-time}, first panel). It is to be noted that the
unjamming of the left orifice occurs only when there is a flow through
the right orifice. Few simulations without the presence of right
orifice showed that the flow through left orifice, once jammed, does
not unjam on its own even over timescales close to that for an entire
simulation run. Similar qualitative behavior is also observed for
other interorifice distances.

\begin{figure}
  \includegraphics[width=0.85\linewidth]{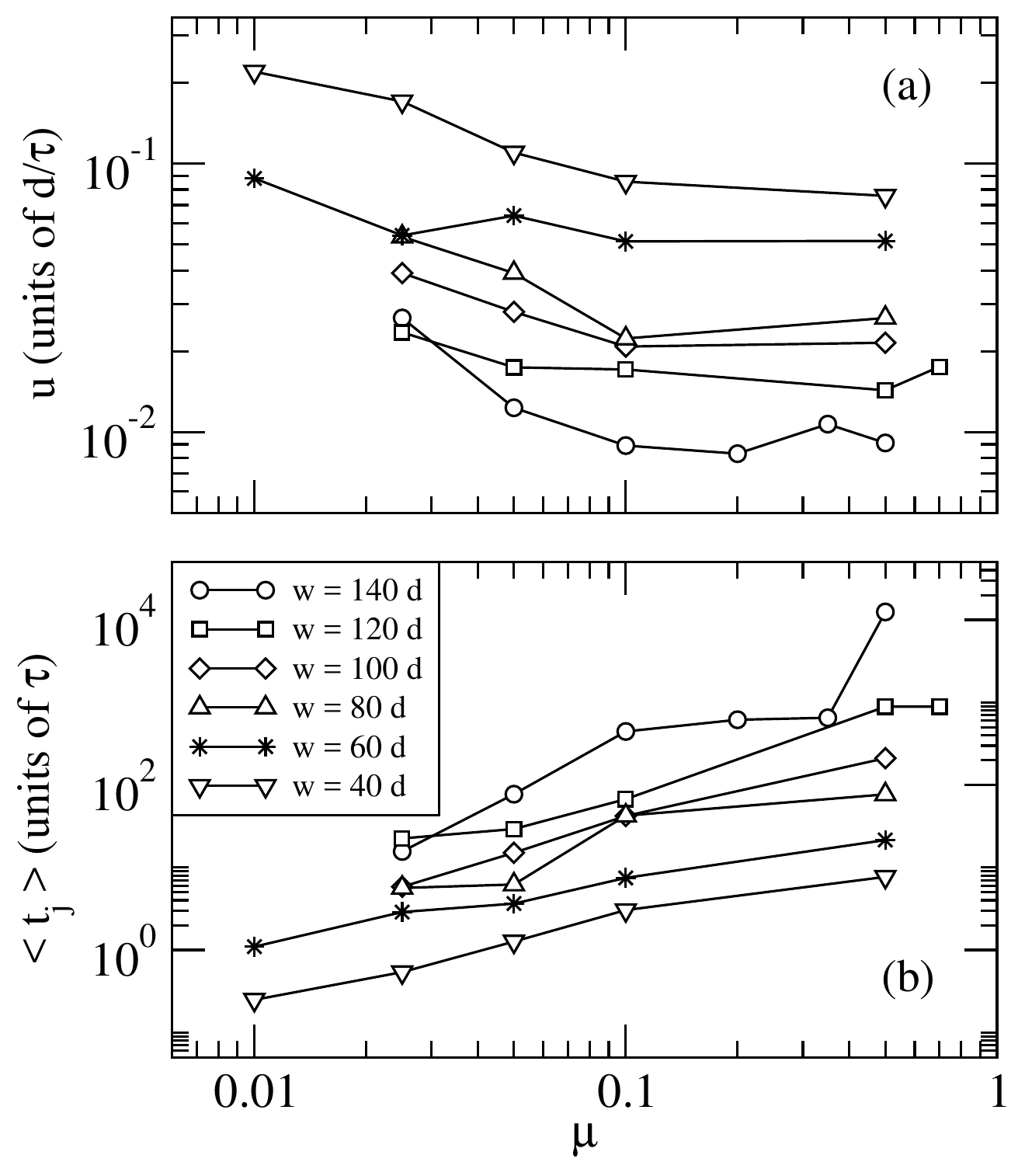}
  \caption{(a) Variation of the rms velocity ($u$) of the flow with
    friction coefficient ($\mu$) for various interorifice distances
    ($w$). The value of $u$ is measured in the white box shown in
    Fig.~\ref{sim-schem} and is obtained as an average over all times
    instances whenever the left orifice remains in the jammed
    state. (b) Variation of mean jammed duration
    $\langle t_{j} \rangle$ with friction coefficient ($\mu$) for
    various interorifice distances ($w$).}
  \label{rmsvel}
\end{figure}

To quantify the effect of fluctuations on unjamming even further, we
have calculated the rms velocity in a larger region
($10d \times 10d$), the same which was used for obtaining the time
dependent velocities shown in Fig.~\ref{flow-time}. As earlier, the
rms velocity is obtained as average over the region as well as over
all durations whenever the left orifice is in the jammed state. The
variation of rms velocity with friction coefficient for various
interorifice distances is shown in Fig.~\ref{rmsvel}(a). The rms
velocity, shows some scatter, but decreases monotonically with
increased friction coefficient and increasing interorifice
distance. Both the trends are expected to arise out of higher
tangential damping during particle-particle contacts, thereby
weakening the fluctuations in the vicinity of the jammed orifice
originating from the continuous flow in the right orifice. A direct
correlation of this effect is observed with the averaged duration of
jammed states ($\langle t_{j} \rangle$) which increases monotonically
with increased values of $\mu$ as well as $w$ as shown in
Fig.~\ref{rmsvel}(b).

The overall effect of the fluctuations on the jamming-unjamming
behavior of the orifice observed over here is somewhat analogous to
that observed previously for single orifice silos. The fluctuations
drivers in friction coefficient and the interorifice distance, then,
correspond to the intensity of vibrations employed for a dry granular
system~\cite{janda09b} or the variation of temperature for a colloidal
system~\cite{hidalgo18} or some random force causing the pedestrians
to exit from a bottleneck~\cite{zuriguel14a}. Such fluctuation driven
flow, also known as nonlocal flow, has been studied previously in
different geometries and under different flow
conditions~\cite{reddy11,nichol12}. It has been shown that the
localised shear gives rise to stress fluctuations leading the material
to yield and flow elsewhere~\cite{reddy11} akin to a self-activated
process. This nonlocal flow has been expressed adequately using
appropriate constitutive equations for the relevant
rheology~\cite{kamrin12}.

\begin{figure}
  \includegraphics[width=1.0\linewidth]{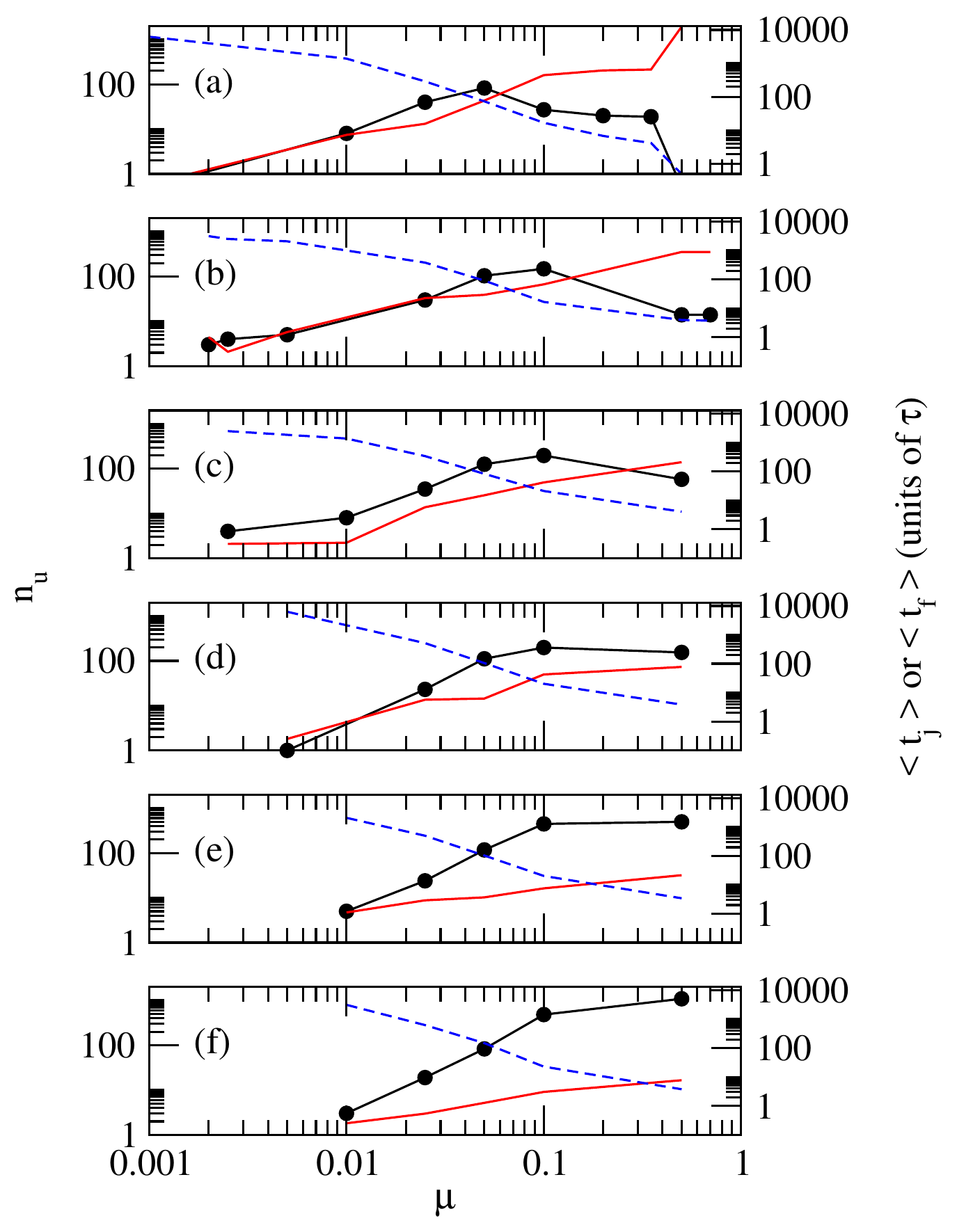}
  \caption{Variation of mean jammed duration $\langle t_{j} \rangle$
    (solid, red lines), mean flow duration $\langle t_{f} \rangle$
    (dashed, blue lines) and the frequency of unjamming $n_{u}$ (lines
    with filled circles) with friction coefficient ($\mu$). The data
    is shown for different interorifice distances, namely, (a)
    $w = 140$d, (b) $w = 120$d, (c) $w = 100$d, (d) $w = 80$d, (e)
    $w = 60$d and (f) $w = 40$d.}
  \label{unjam-freq-time}
\end{figure}

We next discuss the kinematics of jamming and unjamming events. The
duration of every jammed and flow events is, respectively, represented
by $t_{j}$ and $t_{f}$ as mentioned earlier.  The frequency of
unjamming ($n_{u}$) is defined as the number of times the orifice
unjams over the entire simulation time period
($t = 12500 \tau$). The values of $t_{j}$, $t_{f}$, $n_{u}$ are
obtained by monitoring the presence of particles in the outflow from
the left orifice and the averages $\langle . \rangle$ are obtained
over the entire simulation duration.  The variation of $n_{u}$,
$\langle t_{j} \rangle$ and $\langle t_{f} \rangle$ with
friction coefficient ($\mu$) for different interorifice distances ($w$)
employed is shown in Fig.~\ref{unjam-freq-time}.

The effect of friction coefficient ($\mu$) for a fixed value
of $w$ is discussed first followed by the overall dependence on
$w$. Consider the profiles for $w = 140d$ shown in the topmost
panel. Both, $\langle t_{j} \rangle$ shown as red solid lines and
$\langle t_{f} \rangle$ shown as blue dashed lines, show a monotonic
dependence on the friction coefficient, though in opposite
direction. The average time over which the orifice remains jammed,
increases progressively with increased value of $\mu$, while the
average time for which the orifice is flowing,
decreases progressively. As discussed earlier with
respect to Figs.~\ref{velocity-fields} and ~\ref{rmsvel}, the increased
friction leads to 
weaker fluctuations reaching the orifice thereby allowing for longer
duration of the arch (i.e., jammed state) before causing slight
rearrangements leading to unjamming.  The curves for
$\langle t_{j} \rangle$ and $\langle t_{f} \rangle$ cross each other
at some value of crossover friction coefficient, denoted as $\mu_{c}$,
for which the average duration of jammed and flowing states are
identical, for instance, a situation similar to that shown in panel 4 in
Fig.~\ref{flow-time}. 

The frequency of unjamming occurrences, however, shows a nonmonotonic
dependence on the value of $\mu$, with the maximum occurring quite
close to $\mu_{c}$ where the curves for jamming and flowing times
intersect. On either side of the maximum, identical value of frequency
is achievable for two different values of $\mu$, though the origin is
quite opposite to each other. Towards the left side, for smaller
values of $\mu$, the orifice remains in the flowing state for most of
the time with few jamming events, consequently lesser number of
unjamming instances and hence lower $n_{u}$.  The
smaller value of $n_{u}$ on the right side is also the result of
lesser number of unjamming events, but due to the orifice remaining
jammed for a longer duration due to weaker fluctuations reaching the
jammed orifice. The value of $n_{u}$ eventually reaches zero for very
small and very high friction coefficients, which represents,
respectively, a completely jammed state (first panel in
Fig.~\ref{flow-time} and diverging $\langle t_{j} \rangle$) and a
completely flowing state (last panel in Fig.~\ref{flow-time} and
diverging $\langle t_{f} \rangle$).  The friction coefficient (nearly
same as $\mu_{c}$) corresponding to the maximum in the frequency
curves, can then, perhaps, be termed as the flowing-jamming transition
point. The variation in the value of $\mu$, either decreasing below or
increasing above $\mu_{c}$, shifts the system, respectively, towards
either a progressively flowing or a progressively jammed state.

A similar behavior was observed previously by Janda et
al.~\cite{janda09b} in an experimental study on vibrated silo with a
single orifice.  The average jamming time showed a progressive
decrease with an increase in the vibrational intensity. The
progressive, smooth decrease, however was shown to transform to a step
curve around a critical vibrational intensity bifurcating the jammed
and flowing states, if the silo was allowed to flow for an infinitely
long duration of time. The analog of the critical vibrational
intensity in the present case is the crossover friction coefficient
$\mu_{c}$ corresponding to the maximum frequency.

The overall behavior of $\langle t_{j} \rangle$,
$\langle t_{f} \rangle$ and $n_{u}$ is preserved qualitatively for
decreasing interorifice distances as shown in the remaining panels
of Fig.~\ref{unjam-freq-time}, but with quantitative differences. The
value of crossover or transition $\mu_{c}$ increases with decreasing
interorifice distance. This means that for a fixed value of $\mu$,
the jamming dominated regime is obtained for larger interorifice
distance, while flowing dominated regime is obtained at smaller
interorifice distance. For instance, the same fluctuations
originating from the same right orifice for $\mu = 0.2$, are weak
enough to cause unjamming of the other orifice situated $140d$ away,
but are significantly strong enough to cause frequent unjamming of the
second orifice situated only $40d$ away.  The jamming to flowing
transition ($\mu_{c}$), if as defined, does not seem to be unique, but
is a function of the interorifice distance which also serves to
induce independent forcing in the system.  The values of average
jamming and flowing times at crossover or transition point are
significantly reduced at smaller $w$ showing relatively rapid
occurrences of jamming and unjamming instances. The reason for this
being the progressively stronger fluctuations available at the
jamming/unjamming orifice with decreasing values of $w$.  Not
surprisingly, the values of $n_{u}$ become higher at the
crossover friction coefficient with decreasing $w$. While the curves
for $w = 80d$ and lower are incomplete towards higher friction
coefficients, they nevertheless convey the same qualitative
behavior. The relative flattening of the frequency curves at higher
values of $\mu$, but for smaller values of $w$, cannot be commented
due to inadequate data available.

\begin{figure}
  \includegraphics[width=0.9\linewidth]{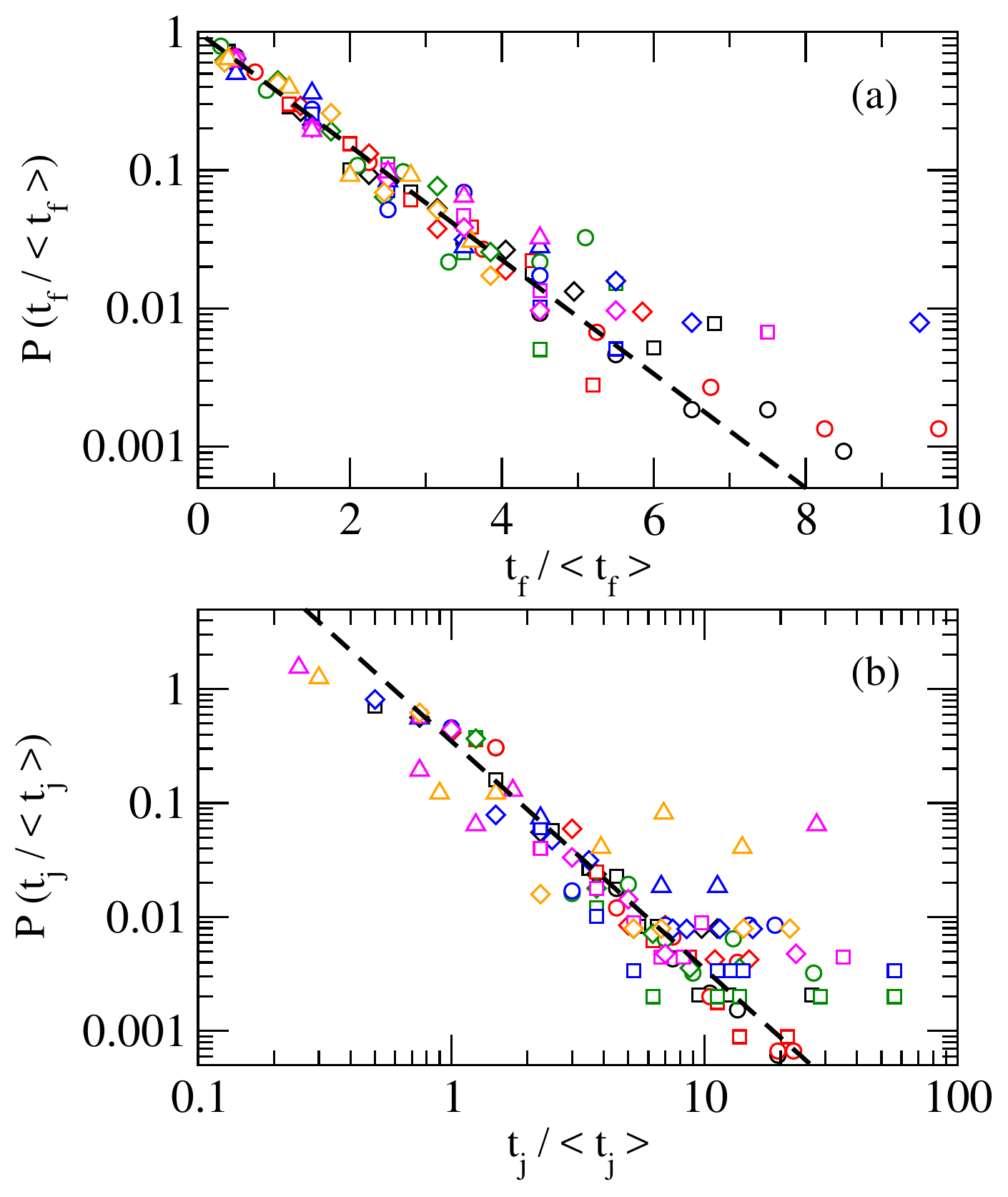}
  \caption{Probability distribution of (a) normalised flow time
    ($t_{f}$) and (b) normalized jammed time ($t_{j}$) obtained for
    various interorifice distances and friction coefficients. Data is
    represented by $\bigcirc$ ($\mu = 0.5$), $\Box$ ($\mu = 0.1$),
    $\Diamond$ ( $\mu = 0.05$) and $\triangle$ ($\mu = 0.025$). The
    color of the symbol represents different values of interorifice
    distances: black ($w = 40d$), red ($w = 60d$), green
    ($w = 80d$), blue ($w = 100d$), magenta ($w = 120d$) and orange
    ($w = 140d$). The dashed line in (a) represents an exponential fit
    while the dashed line in (b) represents a power-law fit with an
    exponent of $2$. See text for more details.}
  \label{distributions}
\end{figure}
 
The distributions of jamming times ($t_{j}$) and flowing times
($t_{f}$), normalised by their respective mean values are shown in
Fig.~\ref{distributions} for four different friction co-efficients and
various interorifice distances. The distributions for both cases seem
to show similar behavior across $\mu$ and $w$ employed, albeit with a
larger scatter in the distributions for the unjamming times as well as
deviations in the tails in few of the cases.  The distribution of the
flowing time shows an exponential behavior [dashed line in
Fig.~\ref{distributions}(a)] for all combinations of $w$ and $\mu$,
except one or two cases.  The exponential behavior is typical of the
random nature of discrete avalanche events and is in accordance with
the behavior observed previously for single~\cite{janda09b,zuriguel11}
as well as multiorifice~\cite{kunte14} silos. The occurrence of
jamming, thus, is not necessarily influenced by the presence of
independent forcing in the form of fluctuations originating from the
second orifice in the system.  The occurrence of unjamming is,
however, clearly dependent on the presence of second orifice and the
fluctuations generated therein. The distributions of jamming times,
thus, do not show an exponential behavior, but seem to exhibit a
power-law behavior with an exponent value of $2$ across the data for
most of the combinations of $w$ and $\mu$ studied. Previous studies
using single orifice silo with varying independent forcing have also
observed power-law tail in the distributions for the unjamming
times~\cite{janda09b,zuriguel14a,zuriguel17}. The exponent value of
two was shown to be closely related to jamming-flowing transition,
with values greater than two dominated by flowing, while those equal
or lower than two dominated by jamming occurrences.

\section{\label{sum}Summary}
The jamming and flowing behavior of granular material exiting through
a narrow orifice is investigated in the presence of another
continuously flowing wide orifice located in the vicinity for varying
interparticle friction coefficients. Intermittent flow, consisting of
sequential jammed and flowing events, is observed to occur through the
smaller orifice.  The mean time duration of jammed events increases
monotonically with increasing friction coefficients, eventually
diverging at very high friction coefficient resulting in a permanently
jammed state. The opposite behavior is observed for the mean time
duration of flowing events which exhibits a permanently flowing state
at small enough friction coefficient. The friction coefficient
manifests itself by influencing the magnitude of the intensity of
fluctuations reaching the narrow orifice arising out of several
interparticle contact interactions in the system leading to an
intermittent flow.

The frequency of the unjammed events ($n_{u}$) exhibits a
nonmonotonic behavior comprising of a gradual increase followed by a
gradual decrease with increasing value of friction coefficient. The
crossover friction coefficient $\mu_{c}$ corresponding to the maximum
in the value of $n_{u}$ can be thought to be as jamming-to-flowing
transition point. A progressive decrease below or increase above $\mu_{c}$,
respectively, shifts the system monotonically towards progressively
increased duration of flowing or jammed events. The value of
$\mu_{c}$ shifts towards higher values for decreasing interorifice
distances accompanied by progressively increasing corresponding
frequency values. The distributions of flowing time durations exhibit
an exponential tail in accordance with a typical randomly occurring
event independent of the induced forcing. The distributions for the
jammed duration, however, show a slower power-law decay and a definite
dependence on the induced forcing.

The interparticle friction coefficient governs the momentum transfer
between contacting particles causing them to move either slowly or
faster. Its variation, in principle, can be considered to represent
varying momentum transfer through different modes of
independent forcing incorporated previously to study unjamming,
ranging from dry granular material~\cite{janda09b} through colloidal
suspensions~\cite{zuriguel17} to motion of self-propelled
vehicles~\cite{patterson17} and living agents~\cite{zuriguel14a}
across narrow constrictions. This inference which is obviously valid
in the absence of any other mechanism governing momentum transfer, for
instance collision in granular system, nevertheless provides a more
generic nature to the observed behavior in this work. More interesting
would be study the effect of friction on jamming-unjamming behavior in
tilted silos~\cite{thomas13,thomas15,thomas16} which provides tilt
angle as another controlling parameter and for more practical cohesive
systems which provides altered lengthscale to account for cluster size
instead of single particle size.

\begin{acknowledgements}
A.V.O. gratefully acknowledges the financial support from Science \&
Engineering Research Board, India (Grant No. SB/S3/CE/017/2015).
\end{acknowledgements}

\bibliography{unjam}

\end{document}